# High-field THz source centered at 2.6 THz

WEI CUI[1], EESWAR KUMAR YALAVARTHI[1], ASWIN VISHNU RADHAN1[1], MOHAMMAD BASHIRPOUR[1], ANGELA GAMOURAS[1,2] AND JEAN-MICHEL MÉNARD[1,2]*

[1]*University of Ottawa, Department of Physics, 25 Templeton Street, Ottawa, Ontario K1N 6N5, Canada*
[2]*National Research Council Canada, 1200 Montreal Road, Ottawa, Ontario K1A 0R6, Canada*
*\*jean-michel.menard@uottawa.ca*

**Abstract:** We demonstrate a table-top high-field terahertz (THz) source based on optical rectification of a collimated near-infrared pulse in gallium phosphide (GaP) to produce peak fields exceeding 300 kV/cm with a spectrum centered at 2.6 THz. The experimental configuration, based on tilted-pulse-front phase matching, is implemented with a phase grating etched directly onto the front surface of the GaP crystal. Although the THz generation efficiency starts showing a saturation onset as the near-infrared pulse energy reaches 0.57 mJ, we can expect our configuration to yield THz peak fields up to 866 kV/cm when a 5 mJ generation NIR pulse is used. This work paves the way towards broadband, high-field THz sources able to access a new class of THz coherent control and nonlinear phenomena driven at frequencies above 2 THz.

## 1. Introduction

High-field terahertz (THz) sources [1] are widely applied to explore nonlinear THz properties of various materials including semiconductors [2], 2D materials [3], and gases [4]. These sources also allow experimentalists to access new regimes of high-harmonic generation [5], free electron acceleration [6], and coherent control [7–9]. Most research groups working on these topics have employed the tilted-pulse-front technique based on an ultrafast near-infrared (NIR) laser to generate high-field THz pulses by optical rectification inside a $LiNbO_3$ nonlinear crystal [10,11]. Generally, This table-top technique provides a THz electric field in the order of several hundreds of kV/cm, with a record value of 6.3 MV/cm [12], featuring a spectrum which peaks around 1 THz and then gradually decreases in amplitude as frequencies approach the phonon resonance of $LiNbO_3$ at 4.5 THz [13,14]. Modifications to this standard configuration have recently led to a more efficient generation of high-field THz pulses with spectral components up to 4 THz [15]. However, 98% of the pulse energy in this configuration is still contained within the frequency range below 2 THz, preventing most high-field applications requiring a spectrum centered at higher frequencies. Cooling the $LiNbO_3$ crystal down to cryogenic temperatures can also extend the spectrum towards higher frequencies by reducing losses due to the phonon absorption tail. For example, this approach has enabled the generation of spectral lines peak at 2.7 THz with a peak field strength around 25 kV/cm, and peak 4 THz, with a peak field strength of about 13 kV/cm with a THz beam size around 1 $mm^2$ [16]. However, the use of cryogenic equipment complicates these experiments. There is therefore a general lack of a straight-forward table-top experimental configuration able to generate high-field THz pulses with a spectrum centered at frequencies exceeding 2 THz. Such a system is crucial for providing experimentalists with access to a new range of phenomena such as phonon-assisted nonlinearities [2,17], coherent control of Bose-Einstein condensation in semiconductor microcavities [18] and saturable transitions in molecular gases [4]. Although other high-field THz generation techniques using air plasma or metallic spintronic emitters have reported peak fields above 1 MV/cm [19–21], the spectral bandwidth of these sources extend from 0.5 to 30 THz. This large bandwidth limits many nonlinear applications as the energy available to address a specific transition at a given frequency remains relatively weak [22]. Finally, organic materials, such as DAST and DSTMS, also display great potential for generating high-field THz up to 30 THz [23–26]. However, organic crystals typically have a damage threshold below 20 mJ/cm$^2$ [25–27], which is at least 60 times lower than that of semiconductor crystals, such as GaP [28,29]. This relatively low damage threshold may be a limiting factor for long-term stability at high incident power.

In this work, we demonstrate a scheme to generate high-field THz pulses centered at 2.6 THz using a phase grating directly etched onto the surface of a nonlinear crystal to enable non-collinear phase-matched tilted-pulse-front THz generation. This technique has been proposed theoretically [30–36] and then demonstrated experimentally [37–39] with surface gratings on ZnTe and LiNbO$_3$ crystals to achieve high THz generation efficiencies at frequencies <2 THz. In comparison to the standard tilted-pulse-front technique in LiNbO$_3$, this configuration eliminates imaging distortions of the diffracted generation pulse, hence improving THz beam quality as well as THz generation efficiency [30,33,37]. Recently, broadband THz generation was demonstrated from 0.1 to 6 THz with a phase grating etched on the surface of a 2 mm thick GaP crystal [39]. Here we use the same material, but we rely instead on a different geometry using a collimated 0.57 mJ NIR pulse to generate a THz pulse with a peak field reaching 303 kV/cm at a central frequency of 2.6 THz. More importantly, there are two indications that our configuration is able to yield much larger THz peak fields: (1) NIR pulse energies to generate THz radiation can be increased by two orders of magnitude before reaching the GaP damage threshold, and (2) we observe a quasi-linear relationship between the incident NIR pulse energy and the emitted THz amplitude, with a slight saturation onset only observed at the highest energies in our experiments.

## 2. Experiments

Fig. 1 shows a diagram of the experiment apparatus for high-field THz generation in a GaP crystal. This crystal is patterned with a surface phase grating to efficiently diffract the incident NIR beam and enable tilted-pulse front phase-matching conditions [10,40]. The optical source is a commercial Yb:KGW regenerative amplifier system generating 265 fs pulses with a center wavelength of 1035 nm, a pulse energy of 1 mJ and a repetition rate of 3 kHz. The output laser beam is focused in air with a 1 m focal length lens. A 1 mm thick BK7 window is placed before the focus to broaden the NIR spectrum from a bandwidth of 3.5 THz to 7.2 THz through self-phase modulation (SPM). The NIR spectra of the pulse before and after the BK7 window are shown in Fig. 1(a). The laser beam is then guided through a set of chirped mirrors, providing a total dispersion of -3000 fs$^2$, to compensate for the positive dispersion in the SPM process and compress the pulse to ~80 fs in the time domain. The autocorrelation traces of the NIR pulses before and after the BK7 window and chirped mirrors are shown in Fig. 1(b).

The NIR beam in the first arm is collimated onto a 1-mm thick 110-oriented GaP semiconductor crystal with a surface-etched phase grating [39,41] to enable efficient THz generation by optical rectification. This NIR beam has a 1/e$^2$ diameter of 3.2 mm and a pulse energy of 0.57 mJ. A grating pitch of $\Lambda = 1.6$ μm, which yields a diffraction angle of 11.7° inside the GaP crystal, enables non-collinear phase-matching conditions leading to broadband THz generation from 0.5 to 6 THz with a geometry allowing both the incident NIR pulses and generated THz pulses to propagate along a direction normal to the crystal plane. In this geometry, the 1$^{st}$ order diffraction angle of the NIR pulse is also equal to its pulse-front-tilt angle [42]:

$$\theta_{tilt} = \cos^{-1} \frac{n_g(\omega_{NIR})}{n(\omega_{4THz})}, \tag{1}$$

where n$_g$(ω$_{NIR}$) is the group index of the NIR generation beam centered at 1035 nm and n(ω$_{4THz}$) is the phase index at 4 THz inside of the generation crystal. The grating filling ratio is 50%, and the target height modulation is 245 nm. This corresponds to an optical π-phase difference between optical light rays passing through the top and bottom sections of the grating, which reduces the $0^{th}$ diffraction order because of destructive interference. In a transmission geometry, we measure 44% of the incident NIR power in both ±1 diffraction orders, while only 0.7% of the incident power remains in the $0^{th}$ diffraction order. Note that the presence of the grating also creates an effective index layer at the air-GaP interface reducing the Fresnel reflection coefficient by about 5% at the front crystal surface [41,42]. The generated THz radiation is collected by an off-axis gold parabolic mirror PM1 in Fig. 1(c) with 1/2" diameter and 1/2" focal length. In the same figure, the subsequent parabolic mirrors, with a 2" diameter and 2" focal length (PM2-PM5), are arranged in a standard terahertz time-domain spectroscopy (THz-TDS) configuration. The gating pulse in the second arm is overlapped with the focused THz transient inside a 0.1 mm-thick 110-oriented GaP detection crystal to resolve the oscillating THz transient with electro-optical sampling (EOS) detection.

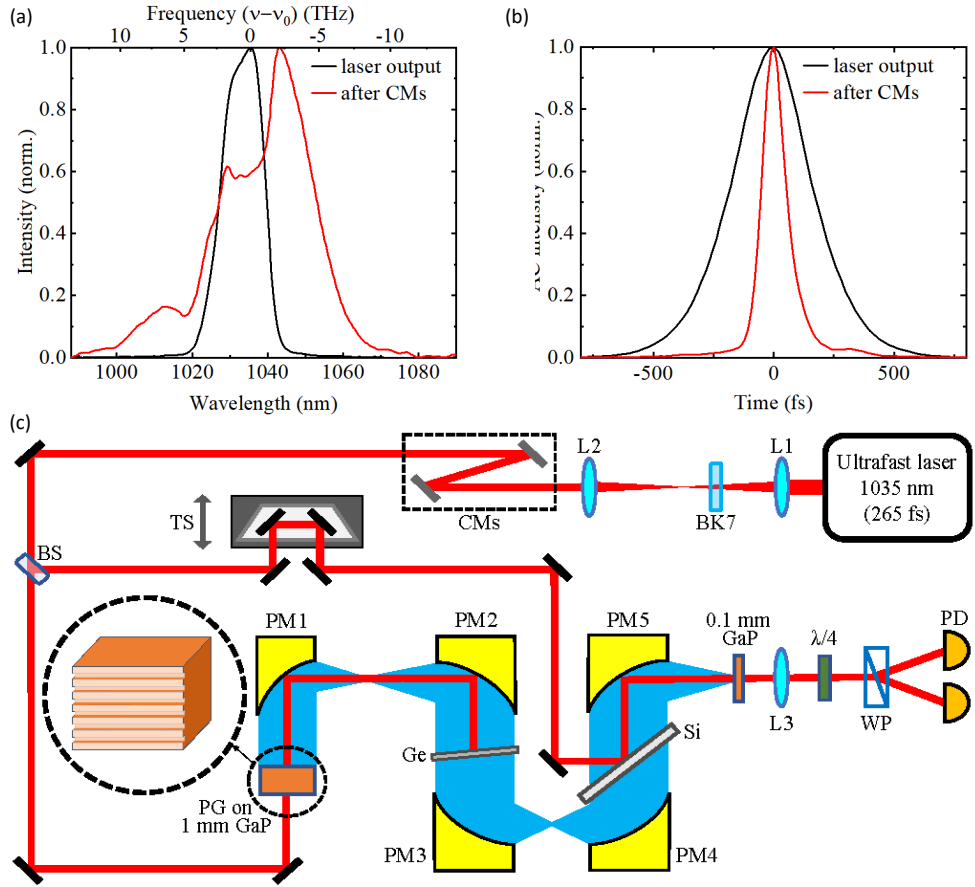

Fig. 1. (a) and (b) Spectra and corresponding autocorrelation traces of the NIR laser pulses measured at the laser output (black line) and after spectral broadening in BK7 and temporal compression with chirped mirrors (CMs) (red line). We observe a spectral broadening from 3.5 THz to 7.2 THz (FWHM) and a reduced pulse duration from 265 fs to 81 fs (FWHM). The reference frequency $\nu_0$ corresponds to the center wavelength of 1035 nm. (c) Schematic of the high-field THz setup. The focused NIR pulses are first passed through a 1-mm thick BK7 window to broaden the spectrum by self-phase modulation. A THz-TDS scheme is then used to generate and detect THz radiation where the NIR generation beam is collimated onto the THz generation crystal. The system is operated in a dry-air purged environment. Optical components to build the setup are labelled above as follows: L1: lens, f = 100 cm; L2: lens, f = 70 cm; CMs: -250 $fs^2$ each; BS: beamsplitter; TS: translation stage; PG on 1 mm GaP: 110-oriented 1 mm-thick GaP crystal with a phase grating on the incident surface; Ge: germanium wafer; Si: silicon wafer; PM: Parabolic mirror; 0.1 mm GaP: 110-oriented 0.1 mm-thick GaP crystal; L3: lens of a focal length of 5 cm; λ/4: quarter-wave plate; WP: Wollaston prism; PD: photodetector.

## 3. Results and Discussion

Fig. 2 shows the measured high-field THz transient and corresponding spectral amplitude, obtained with the Fourier transform. The spectral bandwidth extends up to 6.3 THz with a peak centered at 2.6 THz. The multi-cycle pulse is attributed to group velocity dispersion in the 1 mm-thick GaP generation crystal stretching the pulse in time. We evaluate the electric field strength of the THz pulse at the focus between PM3 and PM4 since this is a practical position to insert a sample in the setup. We evaluate the THz electric field $E_{THz}$ with the following equation [43]:

$$E_{THz} = \frac{A-B}{A+B} \frac{\lambda_{gat}}{2\pi r_{41} n_0^3 L t_{tot}}, \qquad (2)$$

where $A$ and $B$ are the voltages on the PDs, $\lambda_{gat}$ is the central wavelength of the NIR gating beam (1035 nm), $r_{41} = 1$ pm/V is the electro-optical coefficient of GaP at 1035 nm [44], $n_0$ is the refractive index of the GaP at 1035 nm, $L = 0.1$ mm is the thickness of the GaP detection crystal, and $t_{tot}$ is the transmission coefficient taking into account the THz transmission through the GaP detection crystal, the Si wafer after PM4 and the Ge wafer after PM2. Considering this geometry, we obtain a peak field of 303 kV/cm at the focus of PM1.

The dynamic range of our system reaches 80 dB at 2.6 THz and remains above 60 dB between 0.6 THz to 5.3 THz (Fig. 2b). This figure of merit is calculated by dividing the THz spectral intensity by the noise floor, which is measured by blocking the THz beam and then fitted to a model $A*(1/f+B)$, where $f$ is the frequency, and $A$ and $B$ are fitting parameters [45].

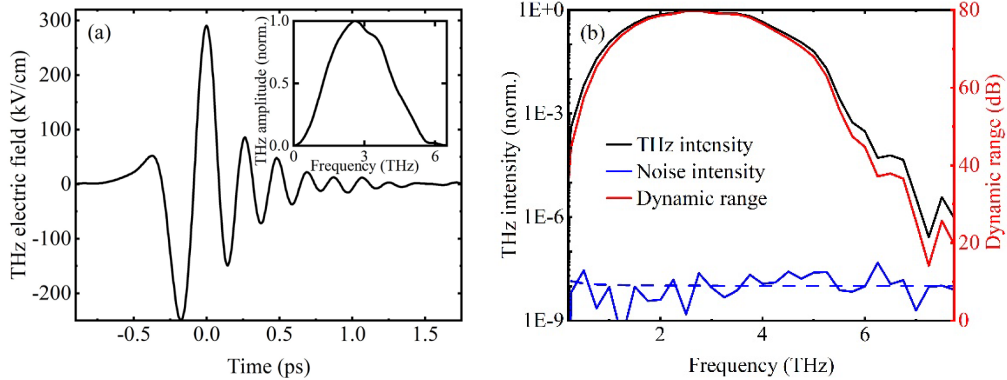

Fig. 2. (a) Time-resolved high-field THz transient. The inset shows the corresponding spectral amplitude calculated with the Fourier transform. (b) THz spectral intensity (black curve) and noise floor (blue curve). The dashed blue line is the noise floor fitted with the model $A*(1/f+B)$, where $f$ is the frequency and both $A$ and $B$ are fitting parameters. The dynamic range (red curve) is calculated based on [45].

To confirm the THz field strength, we also measure the THz power with a thermal detector. A Golay cell, calibrated using a blackbody source, measures a power corresponding to a THz pulse energy of 2.8 nJ immediately after the detection crystal. Considering the Fresnel transmission coefficients of the GaP detection crystal and the Si wafer, the THz pulse energy at the focus between PM3 and PM4 corresponds to 16.8 nJ. Based on the energy contained in the measured time-resolved transient, and the measured THz beam diameter of 317 μm ($1/e^2$), the Golay cell measurement corresponds to a peak field of 446 kV/cm, which is 1.5 times higher than the value measured with the EOS. This discrepancy of the THz pulse energy between the different measurement techniques is well known [46] and has also been observed in other work [43]. We consider that the THz peak field calculated from the EOS measurement is a more reliable value.

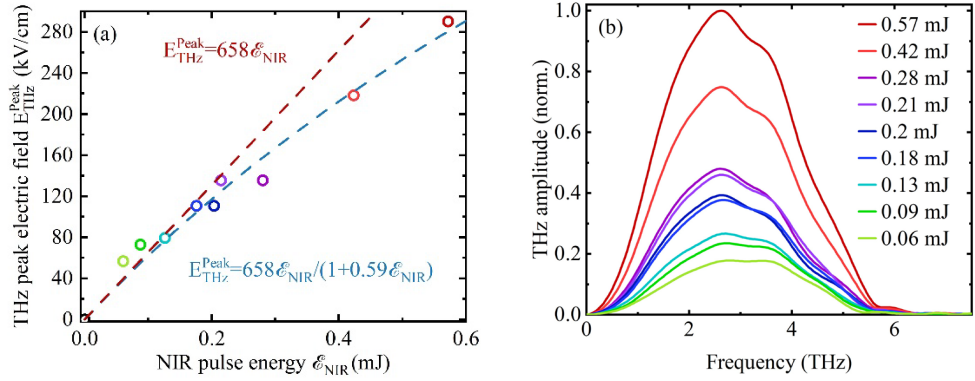

Fig. 3. (a) Measured THz peak electric field $E_{THz}^{Peak}$ strength versus NIR pump pulse energy $\mathscr{E}_{NIR}$. The red dashed line is a linear fit while the blue dashed line corresponds to a saturation model: $a\mathscr{E}_{NIR}/(1+b\mathscr{E}_{NIR})$, where both $a$ and $b$ are fitting parameters. (b) Corresponding THz spectra measured with NIR generation pulses with different pulse energy. We normalize these measurements to the maximum amplitude obtained with a 0.57 mJ NIR generation pulse.

Finally, we look into the possibility to combine our configuration with a more powerful NIR source to achieve even higher THz peak fields. We first determine the crystal damage threshold to the incident NIR pulse, which is a critical parameter to allow the use of higher pulse energies. We perform this test at a wavelength of 1035 nm and a repetition rate of 50 kHz by placing a non-patterned and patterned GaP window at the focus of a 15 cm lens. For both samples, we only notice visible damage on the crystal accompanied by an abrupt drop in the generated THz signal when the incident power exceeds 0.8 W, which corresponds to a peak fluence of 5.6 J/cm$^2$. For comparison purposes, the highest peak fluence used in this work is 14.2 mJ/cm$^2$, which is a factor of 400 lower than this damage threshold. These results indicate that it is indeed possible to significantly increase the incident NIR pulse energy to enable the generation of higher THz peak fields. We also investigate the dependence of the THz peak field $E_{THz}^{Peak}$ as a function of the NIR generation pulse energy $\mathscr{E}_{NIR}$. Fig. 3(a) shows the THz signal as $\mathscr{E}_{NIR}$ is varied from 0.06 mJ to 0.57 mJ. The experimental results are displayed along with two simple models: (1) a linear relationship (red dashed line): $E_{THz}^{Peak} = a\mathscr{E}_{NIR}$, where the slope $a = 658$ kVcm$^{-1}$mJ$^{-1}$ is related to the conversion efficiency, and (2) a modification of the first model (blue dashed curve): $E_{THz}^{Peak} = a\mathscr{E}_{NIR}/(1+b\mathscr{E}_{NIR})$, where $b = 0.59$ mJ$^{-1}$ accounts for saturation effects, which can be caused by two-photon absorption [10]. We observe that the generated THz peak amplitude is linear with $\mathscr{E}_{NIR}$ until 0.2 mJ, which corresponds to a NIR peak fluence of 5.1 mJ/cm$^2$. This saturation onset is 2.5 times higher than the one reported in previous work using the same excitation wavelength, but a non-patterned GaP crystal, and similar to the saturation onset observed with a LiNbO$_3$ crystal in a tilted-pulse front configuration [47]. Fig. 3(b) shows THz spectra generated with different NIR incident pulse energies, which all peak at 2.6 THz. However, we observe a gradual decrease of the THz spectral amplitude around 3.5 THz as $\mathscr{E}_{NIR}$ increases from 0.06 mJ to 0.18 mJ. Similar THz spectral changes at high $\mathscr{E}_{NIR}$ have also been observed previously [48], but further investigation is still required to fully understand this effect. Considering the high damage threshold of GaP and the saturation behavior of $E_{THz}^{Peak}$ observed in our experiments, we predict that a NIR pulse energy of 5 mJ could be used to access a THz peak field of 886 kV/cm with the spectrum centered at 2.6 THz.

## 4. Conclusion

We demonstrate a high-field THz system using collimated NIR pulses impinging on a GaP crystal with a surface phase grating to generate pulses with 303 kV/cm peak field centered at 2.6 THz. This peak field is confirmed by measurements performed with a calibrated Golay cell. We also show that our system operates significantly below the GaP damage threshold and close to a linear regime, allowing more powerful NIR sources to produce even higher peak fields, potentially reaching 1 MV/cm. Also, a laser source at a longer wavelength could be used to reduce the multi-photon absorption processes in GaP, which may help to reduce saturation effects observed in the THz generation process as well as increase the crystal damage threshold. Considering a tilted-pulse-front configuration with an optimal phase grating, this latter approach could, according to numerical models, produce peak fields reaching up to 17 MV/cm at a central frequency of 3 THz [36]. Finally, although our experiment focuses exclusively on GaP as the THz generation crystal, the same optical configuration could be used to generate high-field THz in other materials with a surface phase grating to gain access to different spectral ranges or to increase THz generation efficiencies. This work will pave the way towards a new class of high-field THz sources able to access a spectral range departing from the conventional region below 2 THz and will enable novel nonlinear and coherent control experiments in condensed matter systems.

**Acknowledgement:**

We thank R. Huber for helpful discussions. We acknowledge funding from the National Sciences and Engineering Research Council of Canada (NSERC) (RGPIN-2016-04797) and the NRC-uOttawa Joint Centre for Extreme Photonics.

**References**

1. J. A. Fülöp, S. Tzortzakis, and T. Kampfrath, "Laser-Driven Strong-Field Terahertz Sources," Adv. Opt. Mater. **8**, 1900681 (2020).
2. Y. Lu, Q. Zhang, Q. Wu, Z. Chen, X. Liu, and J. Xu, "Giant enhancement of THz-frequency optical nonlinearity by phonon polariton in ionic crystals," Nat. Commun. **12**, 3183 (2021).
3. I. Katayama, H. Aoki, J. Takeda, H. Shimosato, M. Ashida, R. Kinjo, I. Kawayama, M. Tonouchi, M. Nagai, and K. Tanaka, "Ferroelectric Soft Mode in a $SrTiO_3$ Thin Film Impulsively Driven to the Anharmonic Regime Using Intense Picosecond Terahertz Pulses," Phys. Rev. Lett. **108**, 097401 (2012).
4. P. Rasekh, A. Safari, M. Yildirim, R. Bhardwaj, J.-M. Ménard, K. Dolgaleva, and R. W. Boyd, "Terahertz Nonlinear Spectroscopy of Water Vapor," ACS Photonics **8**, 1683–1688 (2021).
5. B. Cheng, N. Kanda, T. N. Ikeda, T. Matsuda, P. Xia, T. Schumann, S. Stemmer, J. Itatani, N. P. Armitage, and R. Matsunaga, "Efficient Terahertz Harmonic Generation with Coherent Acceleration of Electrons in the Dirac Semimetal $Cd_3As_2$," Phys. Rev. Lett. **124**, 117402 (2020).
6. Z. Tibai, M. Unferdorben, S. Turnár, A. Sharma, J. A. Fülöp, G. Almási, and J. Hebling, "Relativistic electron acceleration by focused THz pulses," J. Phys. B At. Mol. Opt. Phys. **51**, 134004 (2018).
7. S. Schlauderer, C. Lange, S. Baierl, T. Ebnet, C. P. Schmid, D. C. Valovcin, A. K. Zvezdin, A. V. Kimel, R. V. Mikhaylovskiy, and R. Huber, "Temporal and spectral fingerprints of ultrafast all-coherent spin switching," Nature **569**, 383–387 (2019).
8. X. Li, T. Qiu, J. Zhang, E. Baldini, J. Lu, A. M. Rappe, and K. A. Nelson, "Terahertz field–induced ferroelectricity in quantum paraelectric $SrTiO_3$," Science **364**, 1079–1082 (2019).
9. K. Iwaszczuk, M. Zalkovskij, A. C. Strikwerda, and P. U. Jepsen, "Nitrogen plasma formation through terahertz-induced ultrafast electron field emission," Optica **2**, 116–123 (2015).
10. J. Hebling, K.-L. Yeh, M. C. Hoffmann, B. Bartal, and K. A. Nelson, "Generation of high-power terahertz pulses by tilted-pulse-front excitation and their application possibilities," J. Opt. Soc. Am. B **25**, B6 (2008).
11. H. Hirori, A. Doi, F. Blanchard, and K. Tanaka, "Single-cycle terahertz pulses with amplitudes exceeding 1 MV/cm generated by optical rectification in $LiNbO_3$," Appl. Phys. Lett. **98**, 091106 (2011).
12. B. Zhang, Z. Ma, J. Ma, X. Wu, C. Ouyang, D. Kong, T. Hong, X. Wang, P. Yang, L. Chen, Y. Li, and J. Zhang, "1.4-mJ High Energy Terahertz Radiation from Lithium Niobates," Laser Photonics Rev. **15**, 2000295 (2021).

13. N. V. Sidorov, M. N. Palatnikov, V. S. Gorelik, and P. P. Sverbil, "Second-order Raman spectra of a LiNbO$_3$:Tb crystal," Spectrochim. Acta. A. Mol. Biomol. Spectrosc. **266**, 120445 (2022).
14. S. Kojima, "Broadband Terahertz Spectroscopy of Phonon-Polariton Dispersion in Ferroelectrics," Photonics **5**, 55 (2018).
15. L. Guiramand, J. E. Nkeck, X. Ropagnol, X. Ropagnol, T. Ozaki, and F. Blanchard, "Near-optimal intense and powerful terahertz source by optical rectification in lithium niobate crystal," Photonics Res. **10**, 340–346 (2022).
16. J. Hebling, A. G. Stepanov, G. Almási, B. Bartal, and J. Kuhl, "Tunable THz pulse generation by optical rectification of ultrashort laser pulses with tilted pulse fronts," Appl. Phys. B Lasers Opt. **78**, 593–599 (2004).
17. P. Rasekh, M. Saliminabi, M. Yildirim, R. W. Boyd, J.-M. Ménard, and K. Dolgaleva, "Propagation of broadband THz pulses: effects of dispersion, diffraction and time-varying nonlinear refraction," Opt. Express **28**, 3237–3248 (2020).
18. J.-M. Ménard, C. Poellmann, M. Porer, U. Leierseder, E. Galopin, A. Lemaître, A. Amo, J. Bloch, and R. Huber, "Revealing the dark side of a bright exciton–polariton condensate," Nat. Commun. **5**, 4648 (2014).
19. J. Dai, J. Liu, and X.-C. Zhang, "Terahertz Wave Air Photonics: Terahertz Wave Generation and Detection with Laser-Induced Gas Plasma," IEEE J. Sel. Top. Quantum Electron. **17**, 183–190 (2011).
20. A. D. Koulouklidis, C. Gollner, V. Shumakova, V. Y. Fedorov, A. Pugžlys, A. Baltuška, and S. Tzortzakis, "Observation of extremely efficient terahertz generation from mid-infrared two-color laser filaments," Nat. Commun. **11**, 292 (2020).
21. T. Seifert, S. Jaiswal, U. Martens, J. Hannegan, L. Braun, P. Maldonado, F. Freimuth, A. Kronenberg, J. Henrizi, I. Radu, E. Beaurepaire, Y. Mokrousov, P. M. Oppeneer, M. Jourdan, G. Jakob, D. Turchinovich, L. M. Hayden, M. Wolf, M. Münzenberg, M. Kläui, and T. Kampfrath, "Efficient metallic spintronic emitters of ultrabroadband terahertz radiation," Nat. Photonics **10**, 483–488 (2016).
22. H. Y. Hwang, S. Fleischer, N. C. Brandt, B. G. Perkins, M. Liu, K. Fan, A. Sternbach, X. Zhang, R. D. Averitt, and K. A. Nelson, "A review of non-linear terahertz spectroscopy with ultrashort tabletop-laser pulses," J. Mod. Opt. **62**, 1447–1479 (2015).
23. P. Y. Han, M. Tani, F. Pan, and X.-C. Zhang, "Use of the organic crystal DAST for terahertz beam applications," Opt. Lett. **25**, 675–677 (2000).
24. M. Jazbinsek, U. Puc, A. Abina, and A. Zidansek, "Organic Crystals for THz Photonics," Appl. Sci. **9**, 882 (2019).
25. T. O. Buchmann, E. J. Railton Kelleher, M. Jazbinsek, B. Zhou, J.-H. Seok, O.-P. Kwon, F. Rotermund, and P. U. Jepsen, "High-power few-cycle THz generation at MHz repetition rates in an organic crystal," APL Photonics **5**, 106103 (2020).
26. Z. B. Zaccardi, I. C. Tangen, G. A. Valdivia-Berroeta, C. B. Bahr, K. C. Kenney, C. Rader, M. J. Lutz, B. P. Hunter, D. J. Michaelis, and J. A. Johnson, "Enabling high-power, broadband THz generation with 800-nm pump wavelength," Opt. Express **29**, 38084 (2021).
27. B. Monoszlai, C. Vicario, M. Jazbinsek, and C. P. Hauri, "High-energy terahertz pulses from organic crystals: DAST and DSTMS pumped at Ti:sapphire wavelength," Opt. Lett. **38**, 5106–5109 (2013).
28. H. Lee, "Picosecond mid-IR laser induced surface damage on Gallium Phosphate (GaP) and Calcium Fluoride (CaF$_2$)," J. Mech. Sci. Technol. **21**, 1077–1082 (2007).
29. L. P. Gonzalez, S. Guha, and S. Trivedi, "Damage thresholds and nonlinear optical performance of GaP," in *Conference on Lasers and Electro-Optics, 2004. (CLEO)*, Vol. 1, p. 2 pp.
30. L. Pálfalvi, J. A. Fülöp, G. Almási, and J. Hebling, "Novel setups for extremely high power single-cycle terahertz pulse generation by optical rectification," Appl. Phys. Lett. **92**, 171107 (2008).
31. K. Nagashima and A. Kosuge, "Design of Rectangular Transmission Gratings Fabricated in LiNbO$_3$ for High-Power Terahertz-Wave Generation," Jpn. J. Appl. Phys. **49**, 122504 (2010).
32. M. I. Bakunov and S. B. Bodrov, "Terahertz generation with tilted-front laser pulses in a contact-grating scheme," J. Opt. Soc. Am. B **31**, 2549 (2014).
33. Z. Ollmann, J. Hebling, and G. Almási, "Design of a contact grating setup for mJ-energy THz pulse generation by optical rectification," Appl. Phys. B **108**, 821–826 (2012).
34. Z. Ollmann, J. A. Fülöp, J. Hebling, and G. Almási, "Design of a high-energy terahertz pulse source based on ZnTe contact grating," Opt. Commun. **315**, 159–163 (2014).
35. P. S. Nugraha, G. Krizsán, G. Polónyi, M. I. Mechler, J. Hebling, G. Tóth, and J. A. Fülöp, "Efficient semiconductor multicycle terahertz pulse source," J. Phys. B At. Mol. Opt. Phys. **51**, 094007 (2018).
36. Z. Tibai, G. Krizsán, G. Tóth, G. Almási, G. Illés, L. Pálfalvi, and J. Hebling, "Scalable microstructured semiconductor THz pulse sources," Opt. Express **30**, 45246–45258 (2022).
37. J. A. Fülöp, Gy. Polónyi, B. Monoszlai, G. Andriukaitis, T. Balciunas, A. Pugzlys, G. Arthur, A. Baltuska, and J. Hebling, "Highly efficient scalable monolithic semiconductor terahertz pulse source," Optica **3**, 1075 (2016).
38. M. Tsubouchi, K. Nagashima, F. Yoshida, Y. Ochi, and M. Maruyama, "Contact grating device with Fabry–Perot resonator for effective terahertz light generation," Opt. Lett. **39**, 5439 (2014).
39. W. Cui, K. M. Awan, R. Huber, K. Dolgaleva, and J.-M. Ménard, "Broadband and High-Sensitivity Time-Resolved THz System Using Grating-Assisted Tilted-Pulse-Front Phase Matching," Adv. Opt. Mater. **10**, 2101136 (2022).

40. J. Hebling, G. Almási, I. Kozma, and J. Kuhl, "Velocity matching by pulse front tilting for large area THz-pulse generation," Opt. Express **10**, 1161 (2002).
41. M. Bashirpour, W. Cui, A. Gamouras, and J.-M. Ménard, "Scalable Fabrication of Nanogratings on GaP for Efficient Diffraction of Near-Infrared Pulses and Enhanced Terahertz Generation by Optical Rectification," Crystals **12**, 684 (2022).
42. M. Bashirpour, S. Khankalantary, and M. Hajizadeh, "A new hybrid metasurface design for performance improvement of thin film unbiased terahertz photoconductive source," Optik **246**, 167817 (2021).
43. F. Blanchard, L. Razzari, H. C. Bandulet, G. Sharma, R. Morandotti, J. C. Kieffer, T. Ozaki, M. Reid, H. F. Tiedje, H. K. Haugen, and F. A. Hegmann, "Generation of 1.5 µJ single-cycle terahertz pulses by optical rectification from a large aperture ZnTe crystal," Opt. Express **15**, 13212 (2007).
44. M. Nagai, E. Matsubara, M. Ashida, J. Takayanagi, and H. Ohtake, "Generation and Detection of THz Pulses with a Bandwidth Extending beyond 4 THz Using a Subpicosecond Yb-Doped Fiber Laser System," IEEE Trans. Terahertz Sci. Technol. **4**, 440–446 (2014).
45. J. Neu and C. A. Schmuttenmaer, "Tutorial: An introduction to terahertz time domain spectroscopy (THz-TDS)," J. Appl. Phys. **124**, 231101 (2018).
46. E. Castro-Camus, M. Koch, T. Kleine-Ostmann, and A. Steiger, "On the reliability of power measurements in the terahertz band," Commun. Phys. **5**, 1–3 (2022).
47. M. C. Hoffmann, K.-L. Yeh, J. Hebling, and K. A. Nelson, "Efficient terahertz generation by optical rectification at 1035 nm," Opt. Express **15**, 11706 (2007).
48. J. A. Fülöp, Z. Ollmann, C. Lombosi, C. Skrobol, S. Klingebiel, L. Pálfalvi, F. Krausz, S. Karsch, and J. Hebling, "Efficient generation of THz pulses with 0.4 mJ energy," Opt. Express **22**, 20155–20163 (2014).